\documentclass[sigconf]{acmart}
\AtBeginDocument{%
  }

\usepackage{amsmath}
\usepackage{graphicx,acmart-taps}
\usepackage{pdfpages}
\usepackage{xcolor}
\definecolor{Orange}{rgb}{1,0.5,0} 
\newcommand{\revise}[1]{\textrm{\textrm{\textcolor{black}{#1}}}}

\newcommand{\sw}[1]{\textrm{\textrm{\textcolor{black}{#1}}}}

\newcommand{\rev}[1]{\textrm{\textrm{\textcolor{black}{#1}}}}

\usepackage[normalem]{ulem} 

\AtBeginDocument{%
  }

\copyrightyear{2026}
\acmYear{2026}
\setcopyright{cc}
\setcctype{by}
\acmConference[AutomotiveUI '26]{18th International Conference on Automotive User Interfaces and Interactive Vehicular Applications}{September 20--23, 2026}{Gothenburg, Sweden}
\acmBooktitle{18th International Conference on Automotive User Interfaces and Interactive Vehicular Applications (AutomotiveUI '26), September 20--23, 2026, Gothenburg, Sweden}
\acmDOI{10.1145/3828157.3828769}
\acmISBN{979-8-4007-2814-3/2026/09}

\begin{document}

\title{AI Eyes on the Road: Cross-Cultural Perspectives on Traffic Surveillance}

\author{Ziming Wang}
\authornote{Both authors contributed equally to this research.}
\authornote{The author is currently affiliated with the Norwegian University of Science and Technology (NTNU).} 
 \affiliation{%
  \institution{Chalmers University of Technology}
  \city{Gothenburg}
  \country{Sweden}}
\affiliation{%
  \institution{Stanford University}
  \city{Stanford}
  \state{California}
  \country{USA}}
\email{zmg@stanford.edu}
\orcid{0000-0003-0564-8757}

\author{Shiwei Yang}
\authornotemark[1]
\affiliation{%
  \institution{Ghent University}
  \city{Ghent}
  \country{Belgium}}
\affiliation{%
  \institution{University of Gothenburg}
  \city{Gothenburg}
  \country{Sweden}}
\email{shiwei.yang@ugent.be}
\orcid{0000-0003-2627-0802}

\author{Rebecca Currano}
\affiliation{%
  \institution{Stanford University}
  \city{Stanford}
  \state{California}
  \country{USA}}
\email{bcurrano@stanford.edu}
\orcid{0000-0003-2029-4897}

\author{Morten Fjeld}
\affiliation{%
  \institution{Chalmers University of Technology}
  \city{Gothenburg}
  \country{Sweden}}
  \affiliation{%
  \institution{University of Bergen}
  \city{Bergen}
  \country{Norway}}
\email{fjeld@chalmers.se}
\orcid{0000-0002-9562-5147}

\author{David Sirkin}
\affiliation{%
  \institution{Stanford University}
  \city{Stanford}
  \state{California}
  \country{USA}}
\email{sirkin@stanford.edu}
\orcid{0000-0003-0134-6903}

\renewcommand{\shortauthors}{Z. Wang and S. Yang et al.}

\begin{abstract}
AI-powered road surveillance systems are increasingly proposed to monitor infractions such as speeding, phone use, and jaywalking. While these systems promise to enhance safety by discouraging dangerous behaviors, they also raise concerns about privacy, fairness, and potential misuse of personal data. Yet empirical research on how people perceive AI-enhanced monitoring of public spaces remains limited. We conducted an online survey ($N=720$) using a 3×3 factorial design to examine perceptions of three road surveillance modes—conventional, AI-enhanced, and AI-enhanced with public shaming—across China, Europe, and the United States. We measured perceived capability, risk, transparency, and acceptance. Results show that conventional surveillance was most preferred, while public shaming was least preferred across all regions. Chinese respondents, however, expressed significantly higher acceptance of AI-enhanced modes than Europeans or Americans. Our findings highlight the need to account for context, culture, and social norms when considering AI-enhanced monitoring, as these shape trust, comfort, and overall acceptance.
  
\end{abstract}



\begin{teaserfigure}
\center
  \includegraphics[width=0.86\textwidth]{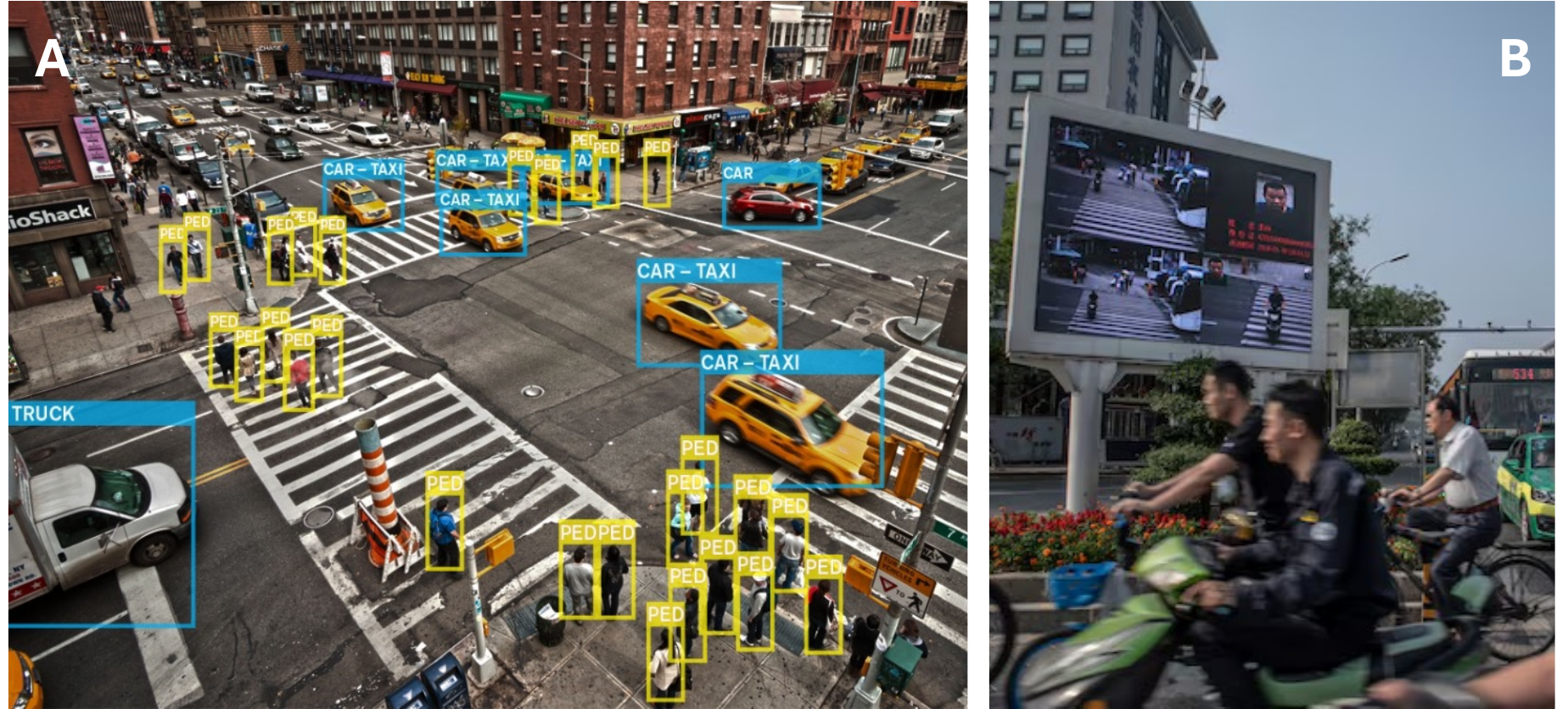}
  \caption{AI-enhanced traffic surveillance systems: (a) AI analyzes traffic conditions and recognizes patterns of vehicles and pedestrians (Image courtesy: Travis Buckner); (b) A roadside screen displays jaywalkers, identified using AI, to deter violations (Image courtesy: Gilles Sabrié).}
  \Description{This image consists of two panels labeled “A” and “B”. Panel A shows an aerial view of a busy urban intersection with several yellow taxis, cars, and pedestrians moving across crosswalks. The image is overlaid with rectangular bounding boxes around various objects, each labeled to indicate what the object is. For instance, pedestrians are labeled as “PED”, cars as “CAR”, taxis as “TAXI”, and a truck is labeled as “TRUCK”. The scene depicts a typical street corner with traffic moving through a four-way intersection. Pedestrians are crossing at different crosswalks while taxis and other vehicles wait or move across the intersection. Panel B shows a large digital billboard or screen on a city street. The screen displays live footage of traffic, which appears similar to the intersection shown in Panel A. In the foreground, several people are riding bicycles and motorbikes along the road in front of the screen. The screen also includes some text, likely related to monitoring traffic or public safety.}
  \label{fig:aits}
\end{teaserfigure}

\maketitle

\section{Introduction}

Vision has long been a cornerstone of intelligence, both in nature and in technology \cite{batchelor1991vision}. In biological evolution, eyes have enabled survival, navigation, and learning. In computing, cameras function as the “eyes” of machines, capturing raw data that computer vision algorithms transform into perception and interpretation. With the rise of artificial intelligence (AI), these “AI eyes” are increasingly deployed on roads worldwide to monitor traffic, enforce regulations, and optimize urban mobility \cite{asu2022argos, thisismoney2024}. Proponents argue that AI-enhanced surveillance can reduce accidents, deter crime, and streamline traffic flow. Yet these promises often come with trade-offs, including loss of privacy, the emergence of the “transparent citizen” \cite{reidenberg20}, and unequal access to digital infrastructures. As AI systems become embedded in everyday life, it is therefore urgent to understand how people perceive and negotiate their presence.

Governance debates further underscore this urgency. The European Union’s General Data Protection Regulation (GDPR) \cite{EU_GDPR} has set global precedents for data privacy, and its newly implemented AI Act is poised to extend such influence to algorithmic governance \cite{feldstein2024evaluating, EUAIAct2024}. Policymakers and experts anticipate ripple effects in the United States and China, shaping not only technical standards but also cultural expectations of surveillance \cite{feldstein2024evaluating}. Against this backdrop, traffic surveillance provides a particularly vivid site for examining public attitudes: it is ubiquitous, consequential for safety, and difficult to opt out of. Unlike consumer technologies, road surveillance systems are woven into the fabric of public life, making their social legitimacy contingent on how citizens experience and evaluate them.

Prior AutomotiveUI \rev{(AutoUI)} research on road safety has largely focused on vehicle-centered solutions (e.g., \cite{tabone2021towardsfuture, clerigo2024safearcross, li2018tocross}), whereas many emerging safety initiatives increasingly rely on public infrastructure such as AI-powered traffic surveillance. This shift from vehicle-centric interaction to infrastructure-based monitoring highlights the need to understand how pedestrians and citizens perceive these systems. \rev{Yet the AutoUI community has already established that ``the interface'' is not confined to the dashboard. A line of work on external human-machine interfaces (eHMIs) shows that pedestrians actively interpret cues broadcast from vehicles and roadside infrastructure, from early studies of explicit versus implicit communication \cite{dey2017pedestrianinteraction} to more recent distance-dependent and scalable designs for multi-pedestrian crossings \cite{dey2020distance, dey2021scalable}. In parallel, a tradition of in-vehicle adaptive HMI research---from early driver-interface design spaces \cite{schmidt2009design} to ambient take-over-request strategies \cite{sadeghian2016assisting} and, more recently, frameworks that gate what information reaches a driver based on estimated mental and perceptual load \cite{gomaa2022mental}---shows that deciding when and how to surface information to a driver is itself a live AutoUI design problem. AI-powered traffic surveillance, which broadcasts judgments to pedestrians (e.g., public shaming displays) and could plausibly surface warnings to drivers via Heads-Up Displays (HUDs) or navigation apps \cite{detjen2022investigating, PCI}, sits within this same design space rather than adjacent to it. What remains open is how such interfaces should be designed across cultural contexts (e.g., \cite{Ge2024-CultureShapesAI, meat-summer}), which is the gap this paper addresses.}

This tension is not abstract. Some countries have experimented with “public shaming” strategies, where surveillance footage of drivers or pedestrians is broadcast to deter misbehavior \cite{drivesmart_2024}. China has gone further by pairing facial recognition with large public displays to expose jaywalkers \cite{Grossman2018ChineseFacial}, see \autoref{fig:aits}(b). Such practices not only amplify privacy concerns but also reveal profound cultural differences in how AI surveillance is understood. While citizens in Europe and the U.S. often emphasize autonomy and control, Chinese respondents have been found to view AI less as a system to constrain and more as one to connect with, even valuing its capacity to influence behavior \cite{Ge2024-CultureShapesAI}. These differences highlight the need for a cross-cultural perspective: the meaning of “AI eyes on the road” cannot be assumed to be universal.

In this paper, we investigate how people in China, Europe, and the USA perceive different modes of traffic surveillance. Our contributions are threefold: (i) we foreground the importance of cross-cultural analysis in debates on AI-enhanced surveillance; (ii) we present findings from an online survey comparing three modes of surveillance—conventional, AI-enhanced, and AI-enhanced with public shaming; and (iii) we propose design recommendations for AI-powered traffic systems that are sensitive to cultural contexts. Together, these contributions advance ongoing conversations about the governance and design of AI, offering insights into how societies might balance safety, privacy, and dignity in the age of algorithmic vision.

\section{Background and Related Work}


\subsection{Road Safety and Public Surveillance}

Road traffic is a socially negotiated environment in which pedestrians and drivers communicate through both implicit and explicit cues. Prior work shows that pedestrians rely on vehicle motion, deceleration, and contextual signals to infer intent, while gestures and eye contact are often used in ambiguous situations \cite{dey2017pedestrianinteraction, ippoliti2025analysisgestures, rao2022investigating, rao2023affectivedriverpedestrian}. These findings highlight that road safety depends on interpreting human behavior rather than purely kinematic information. Several studies have proposed interface-based solutions. For example, external vehicle displays conveying urgency affect pedestrian crossing decisions and promote more cautious behavior \cite{li2018tocross}. Augmented reality approaches further explore pedestrian-facing warnings and virtual traffic signals to assist crossing decisions \cite{tabone2021towardsfuture, clerigo2024safearcross}. Collectively, this line of work focuses on how technology mediates interaction between pedestrians and vehicles. However, these approaches primarily emphasize solutions around vehicles, whereas an increasing number of road safety initiatives rely on public infrastructure.

Urbanization and the rise of smart cities have brought surveillance technologies into sharper focus. Smart city infrastructures increasingly employ interconnected IoT and AI systems \cite{rahman2020scalable, sharma2024video} to monitor traffic conditions and enhance public safety. In this emerging “smart video surveillance paradigm” \cite{rahman2020scalable}, cameras move beyond passive recording to automatically interpret road behavior and support interventions. Such systems promise benefits including accident reduction, improved traffic management, and enhanced emergency response \cite{Adewopo2023, Ashby2017}, but they also raise concerns related to privacy, fairness, and civil liberties \cite{hampapur2003smart}.

HCI research has been central to uncovering the human side of this tension. Studies consistently reveal that coupling cameras with AI intensifies people’s concerns about data collection, storage, and misuse, and may alter behavior when individuals are aware of being monitored (e.g., \cite{chi24drone, lewis2008fears, ONHCR_2022, Guariglia_2023, naturalsoundandproxemics, inaflap}). Although privacy-preserving techniques such as pedestrian tracking or video-to-text reporting have been proposed \cite{Zhang_2010, abdelrahman_2024}, the effectiveness of these approaches ultimately depends on alignment with human values and expectations \cite{Nam2011SmartCity, Pokric_2015, kashef2021smart}.

Importantly, both road behavior and attitudes toward surveillance vary across contexts. A large-scale analysis of driving videos from 233 countries and territories revealed substantial differences in pedestrian crossing timing and negotiation behavior, suggesting that interpretations of “safe” behavior are culturally dependent \cite{Auto25PedestrianPlanet}. Similarly, empirical studies show that perceptions of surveillance vary across contexts. Residents in Detroit adopted a pragmatic “better than nothing” stance \cite{Lu2024}, while European attitudes differed by gender and situational factors, with women generally more accepting despite privacy concerns \cite{golda2022perception, messick2023impact}. In retail settings, acceptance increased when surveillance was transparent and voluntary \cite{brooksbank2022store}, and educational stakeholders in Vietnam viewed surveillance as limited but potentially deterrent \cite{tran2022stakeholders}. Together, these findings suggest that acceptance depends on contextual factors such as perceived benefits, transparency, and cultural norms.

To systematize these factors, the Technology Acceptance Model for Video Surveillance (TAM-VS) identifies perceived usefulness, perceived risk, and system transparency as key determinants of acceptance \cite{TAM_VS}. In the context of AI-powered traffic surveillance, these determinants intersect with culturally shaped traffic behavior and expectations about public monitoring. Understanding pedestrian perspectives within this socio-technical landscape is therefore essential for designing surveillance systems that enhance road safety while remaining socially acceptable across cultural contexts.

\subsection{Human-AI Interaction in Daily Lives}

Human-Centered AI extends beyond designing systems for individual use—it envisions AI as benefiting communities and societies. Human-AI Interaction (HAI) has therefore become increasingly important as AI systems embed themselves in everyday infrastructures. Research documents applications across healthcare \cite{publicai_health1, publicai_health2}, transportation \cite{publicai_trans}, welfare \cite{publicai_welfare}, and public administration \cite{publicai_pa1, publicai_pa2}. From generative chatbots \cite{generative.ai} and recommender systems \cite{ai.suggestion} to autonomous vehicles \cite{autonomous.car} and advanced medical diagnostics \cite{ai.recommendation}, AI technologies are now deeply integrated into social and organizational life. This growing presence has prompted calls for closer integration between HCI and human factors to better understand human responses to interactive AI systems \cite{Iglar_2024}.

While AI offers efficiency and convenience, it also raises concerns about privacy, ethics, and human control \cite{hai}. Acceptance depends not only on technical performance but also on psychological and social factors, including perception, trust, and emotional response \cite{lee2024, guidelines_hci, Gupta2012}. Trust, in particular, is domain-specific: people may prefer human decision-making in healthcare while favoring AI in contexts perceived as objective or data-driven \cite{trust.healthcare, trust.ai}. Identity disclosure further shapes responses, as users evaluate AI differently depending on whether its role is explicit or implicit \cite{Nazaretsky2024}. These findings highlight that acceptance of AI is not only about technical accuracy but also about how AI is socially framed and culturally contextualized.

Surveillance represents a particularly complex domain for HAI because it sits at the intersection of collective benefit and individual rights \cite{Li_2024}. AI-enhanced surveillance can improve efficiency in monitoring road activity, detecting violations, and supporting traffic management, potentially reducing accidents and improving safety. Yet despite this recognition, little research has systematically examined international differences in perceptions of traffic surveillance. This study seeks to address that gap by analyzing cross-cultural perspectives on AI-enhanced road monitoring, where questions of privacy, safety, and trust intersect in everyday public life.

\section{Methodology}

\subsection{Study Design and Procedure} 
Our study employed a mixed 3 × 3 factorial design, with a between-subjects factor of \textbf{region} and a within-subjects factor of \textbf{surveillance mode} as the two independent variables (IVs). The regions considered were China, Europe, and the USA. Here, Europe includes the European Economic Area countries plus Switzerland and the UK. The modes of traffic surveillance included: Conventional Surveillance (CS), AI-Enhanced Surveillance (AS), and AI-Enhanced Surveillance with Public Shaming (PS). The dependent variables (DVs) were four measures of participants' perspectives towards traffic surveillance. See \autoref{table:ariables} for a summary.

\begin{table*}[h]
\caption{Summary of study variables.}
\label{table:ariables}
\begin{tabular}{ p{0.2\textwidth} p{0.25\textwidth} p{0.45\textwidth} } 
\toprule
\textbf{IV1 - Region} & \textbf{IV2 - Surveillance Mode} & \textbf{DV - Measure} \\ 
\midrule
\{China, Europe, USA\} & \{CS, AS, PS\} & Perceived \{Ability, Risk, Transparency\}, Acceptance \\ 
\bottomrule
\end{tabular}
\end{table*}

Participants were asked to rate all three modes of traffic surveillance using the same set of scale items through an online questionnaire. For each mode \revise{(described in \autoref{modes}), participants were presented with a picture as visual stimuli and a textual introduction, followed by a set of Likert-scale questions designed to measure four aspects of perception (detailed in \autoref{measures}). \sw{The display sequence of the modes in the questionnaire was counterbalanced.} The order of the Likert-scale questions within each mode was randomized to reduce potential response bias.} Background questions were asked about participants' gender, age, and the country where they had lived the longest in the past decade. Participants were categorized into three regions (China, Europe, and the USA) based on the country they indicated. The estimated time for completion was 5 to 10 minutes. The online questionnaire was offered in both English and Chinese versions. Participants were recruited through various methods, including snowball sampling and online platforms. Only non-identifiable information was collected. The study was approved by the Institutional Review Board of \rev{Stanford University}.

\subsection{Surveillance Modes} \label{modes}
The distinction between the three modes of surveillance lies in the way data is analyzed and shared, rather than in the type or storage of the data. All three systems consist of numerous cameras streaming pictures and videos, with multiple monitoring sources displayed simultaneously either to traffic \revise{surveillance authorities} or on large public screens. The captured video footage is stored on digital servers for a certain period and can be accessed and reviewed later as needed. The \revise{description} of each system is as follows\rev{, and the stimuli materials used in the study can be found in Supplemental Material in the ACM Digital Library.}

\begin{itemize}
  \item \textbf{Conventional Traffic Surveillance (CS)}: Conventional surveillance systems require human operators to manually analyze the data. One or more operators are responsible for monitoring a specific area. If the operators detect a traffic accident, they notify the relevant personnel for handling.
  
  \item \textbf{AI-Enhanced Traffic Surveillance (AS)}: In this system, AI replaces the manual task of analyzing data found in conventional systems. AI automatically analyzes the captured footage, recognizes traffic accidents and rule violations, and identifies relevant information about vehicles and pedestrians (see \autoref{fig:aits}(a)). Security officers then act based on the information provided by the AI.

  \item \textbf{AI-Enhanced Traffic Surveillance with Public Shaming (PS)}: This application extends the capabilities of AI by not only recognizing traffic rule violations and identifying relevant personal information through facial recognition technology, but also displaying personal details (e.g., facial photos, names, and IDs) of violators on roadside public screens as a deterrent to future violations (see \autoref{fig:aits}(b)).

\end{itemize}

\subsection{Measures} \label{measures} 
Participants rated their perspectives of each mode of presented traffic surveillance across four blocks of questions on a 7-point Likert scale ranging from ``Strongly disagree'' to ``Strongly agree''. Each block had four question items composing a measure.
The four measures were \revise{selected based on TAM-VS developed by~\citet{TAM_VS}}: Perceived Capability, Perceived Risk, Perceived Transparency, and Acceptance. The question items were \revise{also based on the TAM-VS questionnaire~\cite{TAM_VS}}, which was modified to suit traffic surveillance systems. \rev{Some question items, namely Acceptance Items 3 and 4, were reverse-coded because they were negatively worded, ensuring that higher scores consistently represented greater acceptance.} The four items used for each measure are listed below:

\begin{itemize}
  \item \textbf{Perceived Capability (PC)}: The perceived usefulness and reliability of the system in ensuring traffic safety in the observed areas. This includes items, namely, ``reduce traffic accidents'' (PC1), ``improve traffic conditions'' (PC2), ``increase safety'' (PC3), and ``reliability'' (PC4).

  \item \textbf{Perceived Risk (PR)}: The perceived risk associated with the system’s collection and usage of data. This includes concerns about being disadvantaged by ``the processing of the data'' (PR1), ``the breach of the data'' (PR2), ``errors in data collection and processing'' (PR3), and ``the improper use of the data'' (PR4).

  \item \textbf{Perceived Transparency (PT)}: The perceived transparency of the system, including knowing ``the purpose of the system'' (PT1), ``the type of data collected'' (PT2), ``how the data will be processed'' (PT3), and ``who is responsible for the system'' (PT4).

  \item \textbf{Acceptance (AC)}: The overall acceptance of the system, measured by four items: ``I like this system'' (AC1), ``More systems like this should be used'' (AC2), ``Such systems should be illegal'' (AC3), and ``I don’t want this system in my city'' (AC4).
\end{itemize}

\subsection{Sample and Measure Validity}
A total of \sw{720} responses were received. After the initial data screening, 28 unfinished or withdrawn responses were excluded. Additionally, 15 responses from participants residing outside of China, Europe, or the USA were excluded. Furthermore, 76 responses were excluded because the time spent answering the survey was less than 2.5 minutes, which was considered inadequate for a thorough response. Consequently, \sw{601 valid responses were included in the analysis (Europe: n=201; China: n=197; USA: n=203)}. \autoref{fig:age} and \autoref{fig:gender} present the age and gender distributions by region, providing an overview of the sample’s demographics. As the distributions are uneven across regions, we account for these effects in the statistical analysis (see \autoref{statisAnaly}).

The reliability of the questionnaire was initially tested with the following coefficients: Confidence Intervals (CI), Cronbach’s Alpha, and Guttman’s Lambda-6. CI for reliability refers to the range of values within which the true reliability of a measurement instrument is likely to fall. Cronbach’s alpha assesses the extent to which items within a scale are correlated with one another, indicating the internal consistency of the scale~\cite{schmitt1996uses}. Guttman's lambda-6 assesses the extent to which items in a scale can be summed or ordered to reflect a single underlying dimension. Cronbach’s alpha assumes equal weighting for all items, while Guttman’s Lambda-6 does not~\cite{guttman1945basis}. These coefficients were calculated using \texttt{alpha()} from the \texttt{psych} package in R~\cite{R_about}. In this study, all coefficients were within the acceptable range (see \autoref{table:reliability_values} in Appendix for details), indicating good reliability of the scales.

\section{Results}

\subsection{Statistical Analyses and Overview}\label{statisAnaly}
Statistical analyses were performed through R version 4.5.1~\cite{R_about}. \sw{To analyze the effects of Mode and Region on measures, linear mixed-effects models (LMEMs) were conducted using the \texttt{lme4} package\cite{lme_package}. The models included Mode as a within-subjects factor, Region as a between-subjects factor, and their interaction term as fixed effects. Gender and Age were included as covariates to control for their potential influence. P-values for the fixed effects were obtained using Satterthwaite's method for degrees of freedom approximation, as implemented in the \texttt{lmerTest} package\cite{lmerTest_package}. Post-hoc comparisons were performed using estimated marginal means with a Tukey adjustment for multiple comparisons via the \texttt{emmeans} package\cite{emmeans_package}. An alpha level of .05 was used for all statistical tests. We decided to report partial eta squared as the estimate of effect size of fixed effects, denoted as $\eta^2_P$, which is interpreted as small effect size (0.01), medium effect size (0.06), or large effect size (0.14)~\cite{partialeta}; and Cohen’s d as a measure of effect size of simple effects multiple comparisons, denoted as d, which is interpreted as small effect size (0.2), medium effect size (0.5), or large effect size (0.8)~\cite{cohen's}. Partial eta squared and Cohen's d were calculated via the \texttt{effectsize} package\cite{effectsize_package}.}

\aptLtoX{\begin{figure*}[hbt!]

  \centering
    \includegraphics[width=\linewidth]{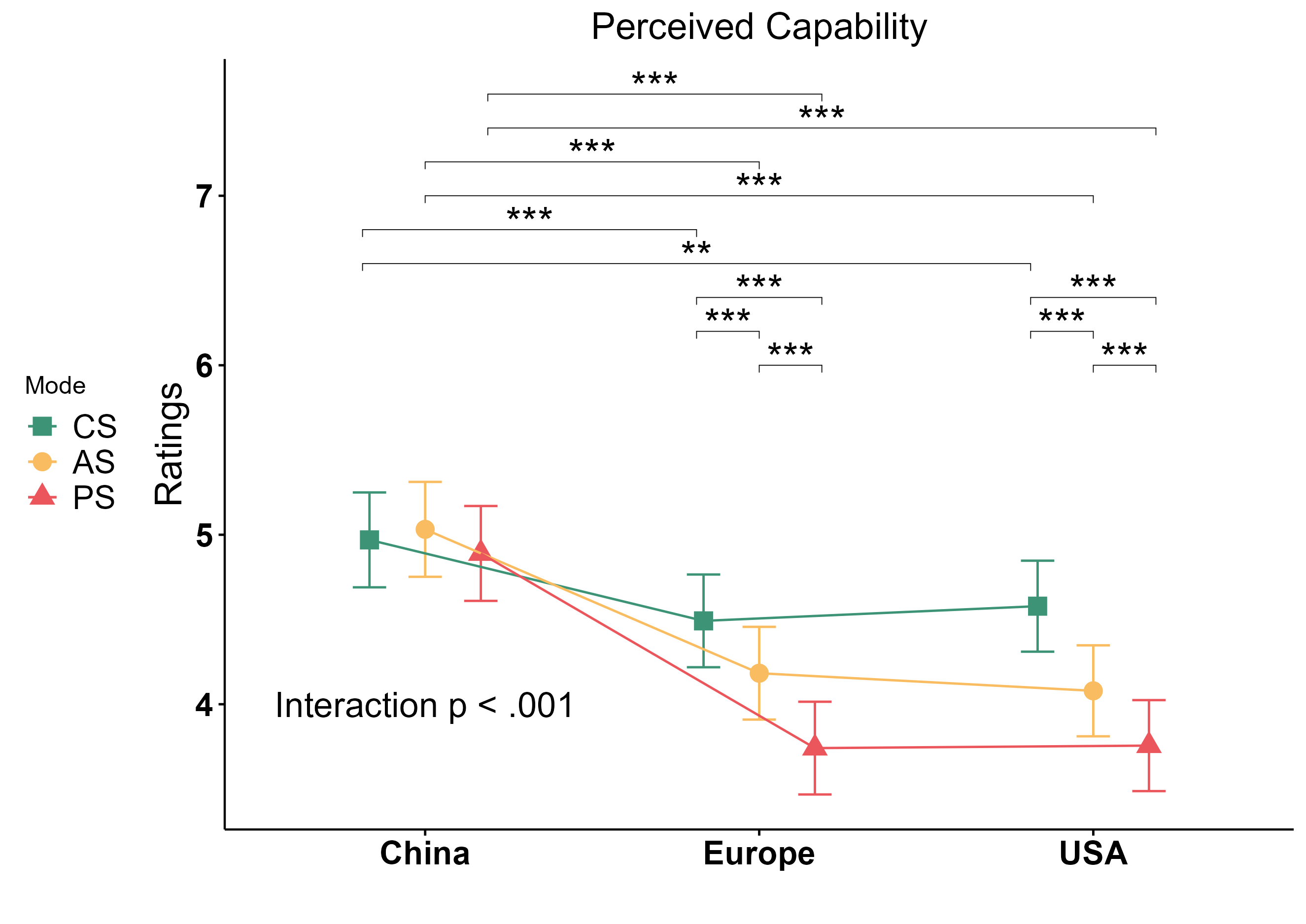} 
    \caption{Interaction plot of Perceived Capability.} 
    \label{fig:PA}
\end{figure*}
\quad
\begin{figure*}
  \centering
    \includegraphics[width=\linewidth]{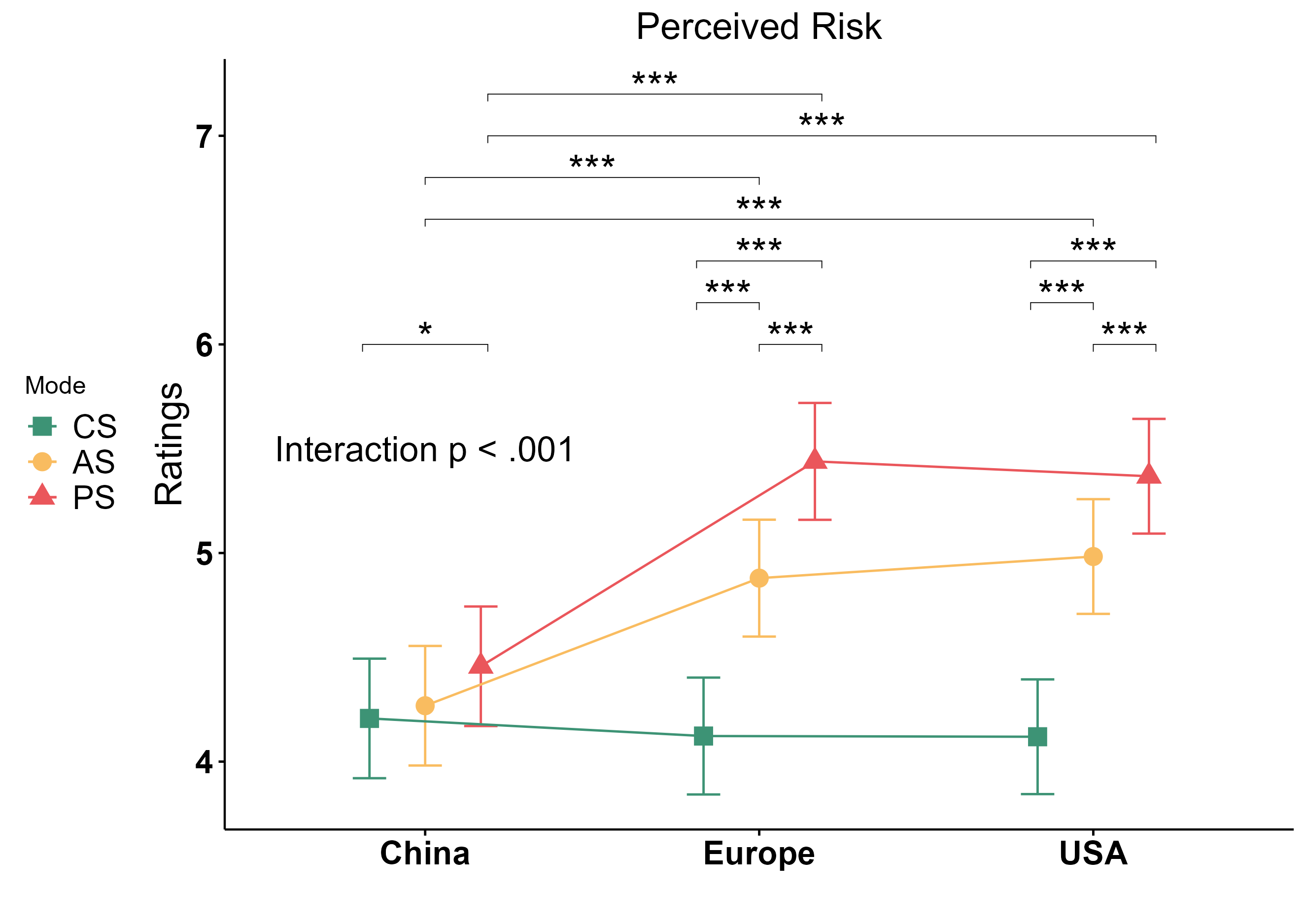} 
    \caption{Interaction plot of Perceived Risk.} 
    \label{fig:PR}
\end{figure*}
\bigskip
\\\textbf{Figures notes:}\\(CS = conventional surveillance; AS = AI-enhanced surveillance; PS = AI-enhanced surveillance with public shaming.)\\All with error bars (95\% confidence interval); \\Simple effects are indicated with asterisks: * for p < .05, ** for p < .01, and *** for p < .001. \\
\bigskip

\begin{figure*}
  \centering
  \includegraphics[width=\linewidth]{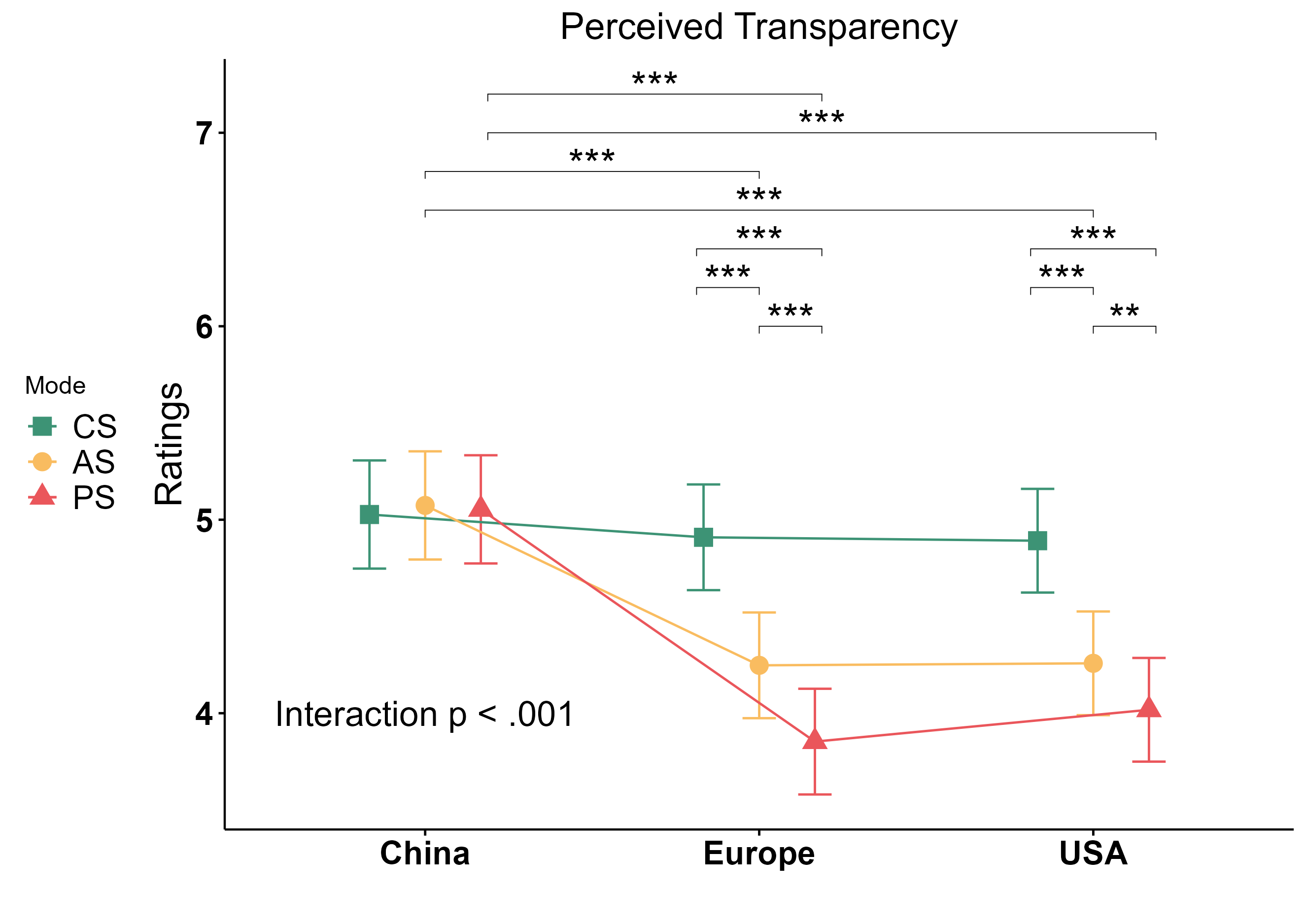} 
    \caption{Interaction plot of Perceived Transparency.} 
    \label{fig:TR}
\end{figure*}
\quad
\begin{figure*}
  \centering
  \includegraphics[width=\linewidth]{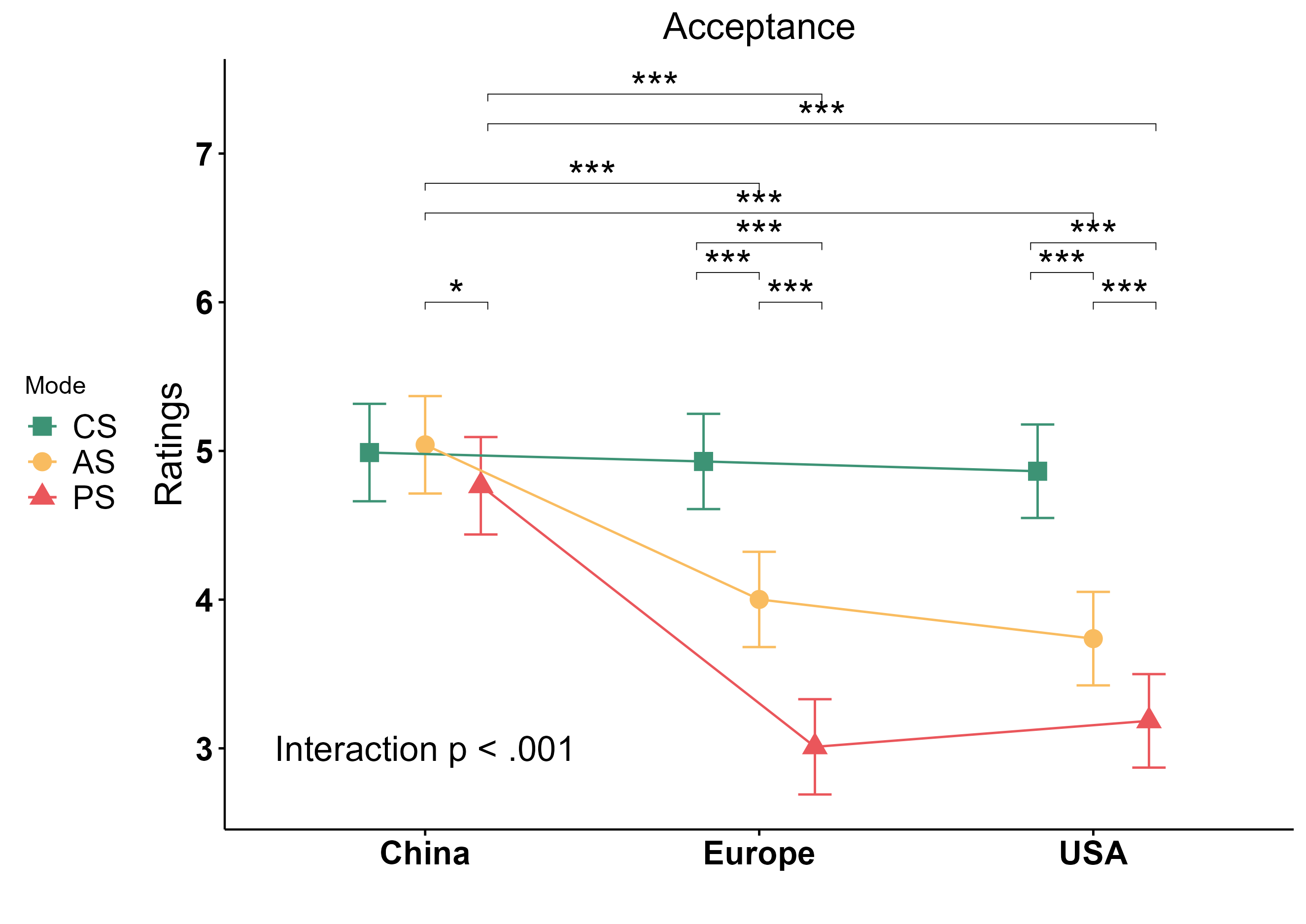} 
    \caption{Interaction plot of Acceptance.} 
    \label{fig:AC}

\end{figure*}}{\begin{figure*}[hbt!]
\centering

\begin{minipage}{.45\linewidth}
  \centering
    \includegraphics[width=\linewidth]{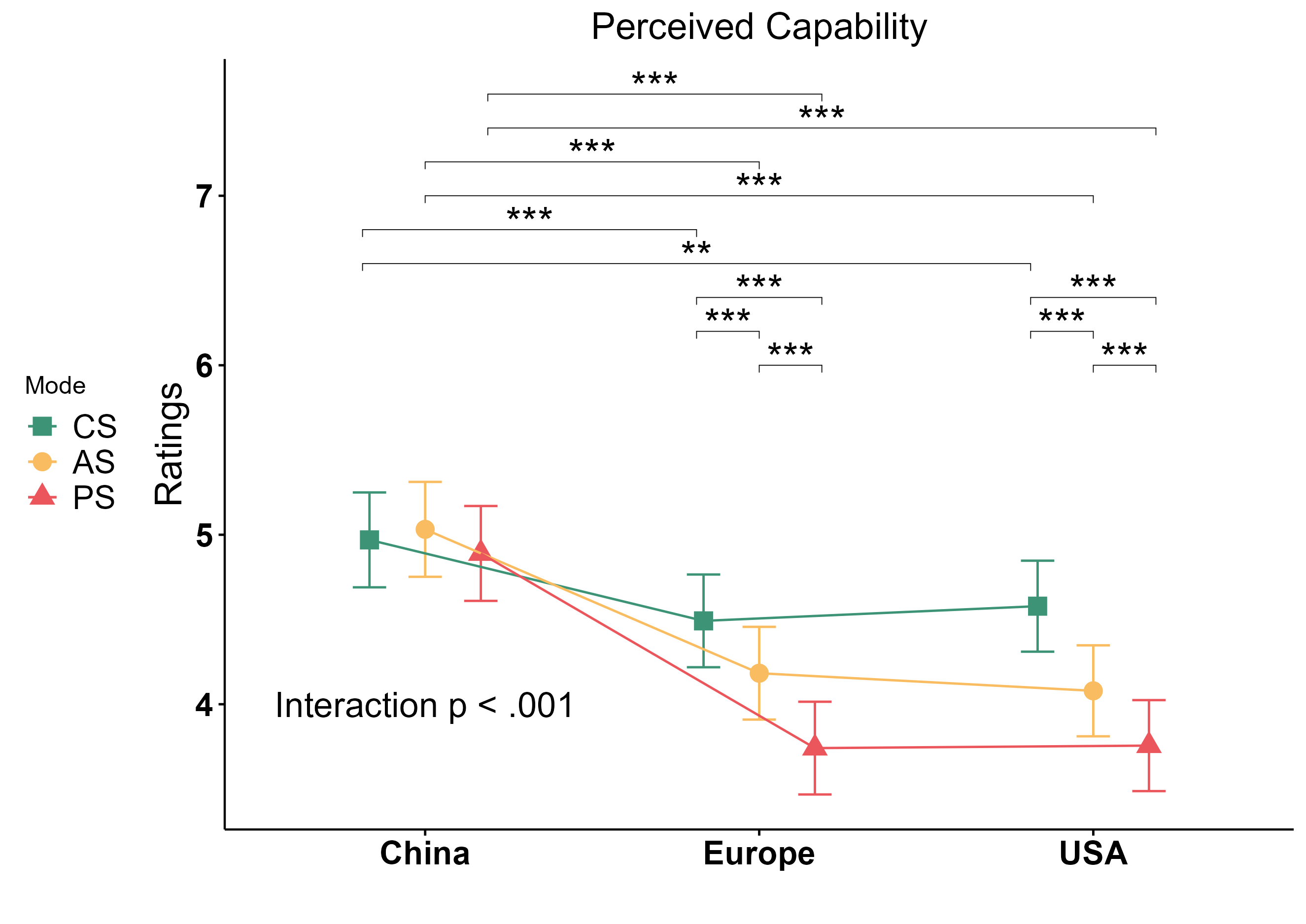} 
    \caption{Interaction plot of Perceived Capability.} 
    \label{fig:PA}
\end{minipage}
\quad
\begin{minipage}{.45\linewidth}
  \centering
    \includegraphics[width=\linewidth]{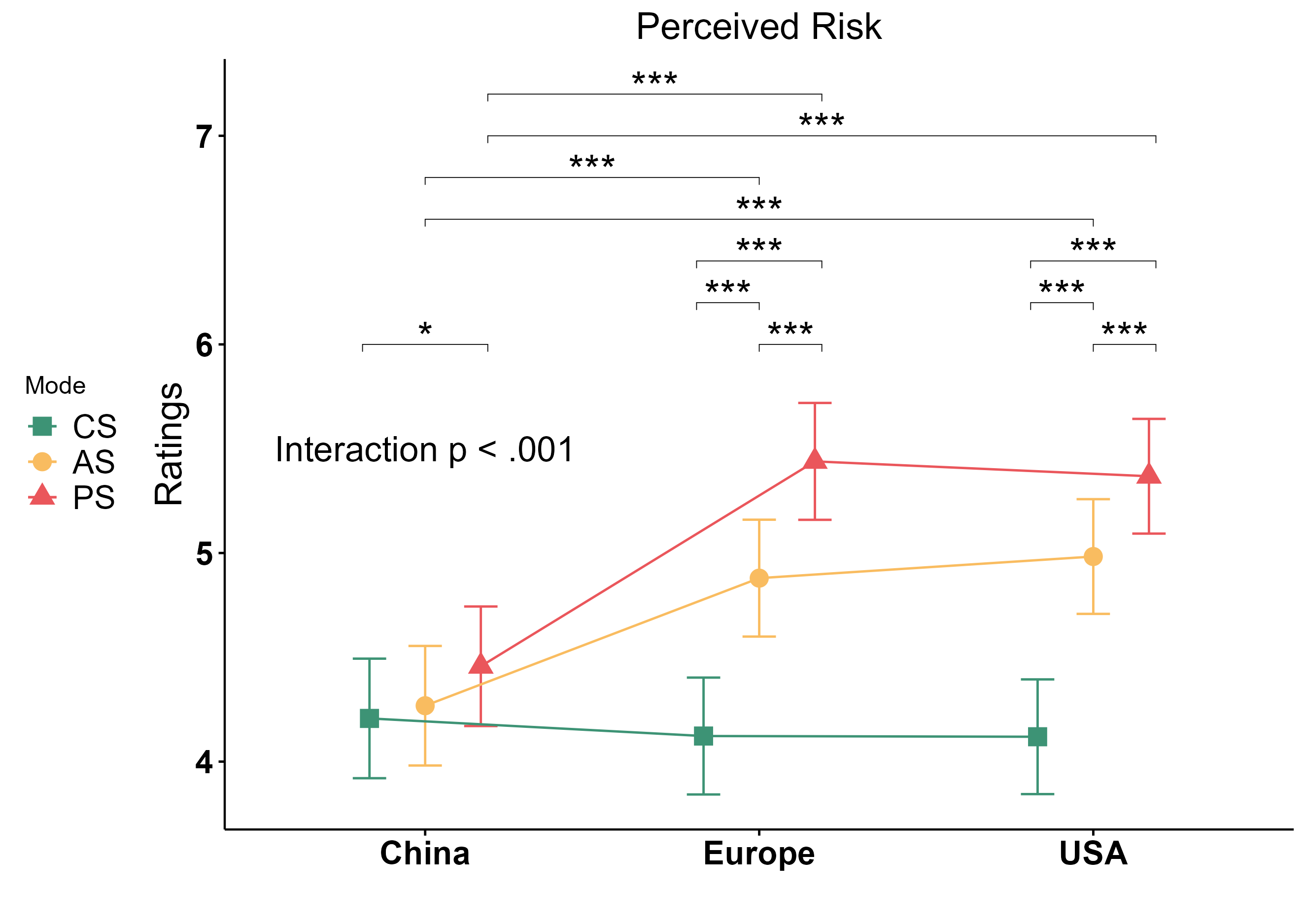} 
    \caption{Interaction plot of Perceived Risk.} 
    \label{fig:PR}
\end{minipage}
\bigskip
\\\textbf{Figures notes:}\\(CS = conventional surveillance; AS = AI-enhanced surveillance; PS = AI-enhanced surveillance with public shaming.)\\All with error bars (95\% confidence interval); \\Simple effects are indicated with asterisks: * for p < .05, ** for p < .01, and *** for p < .001. \\
\bigskip

\begin{minipage}{.45\linewidth}
  \centering
  \includegraphics[width=\linewidth]{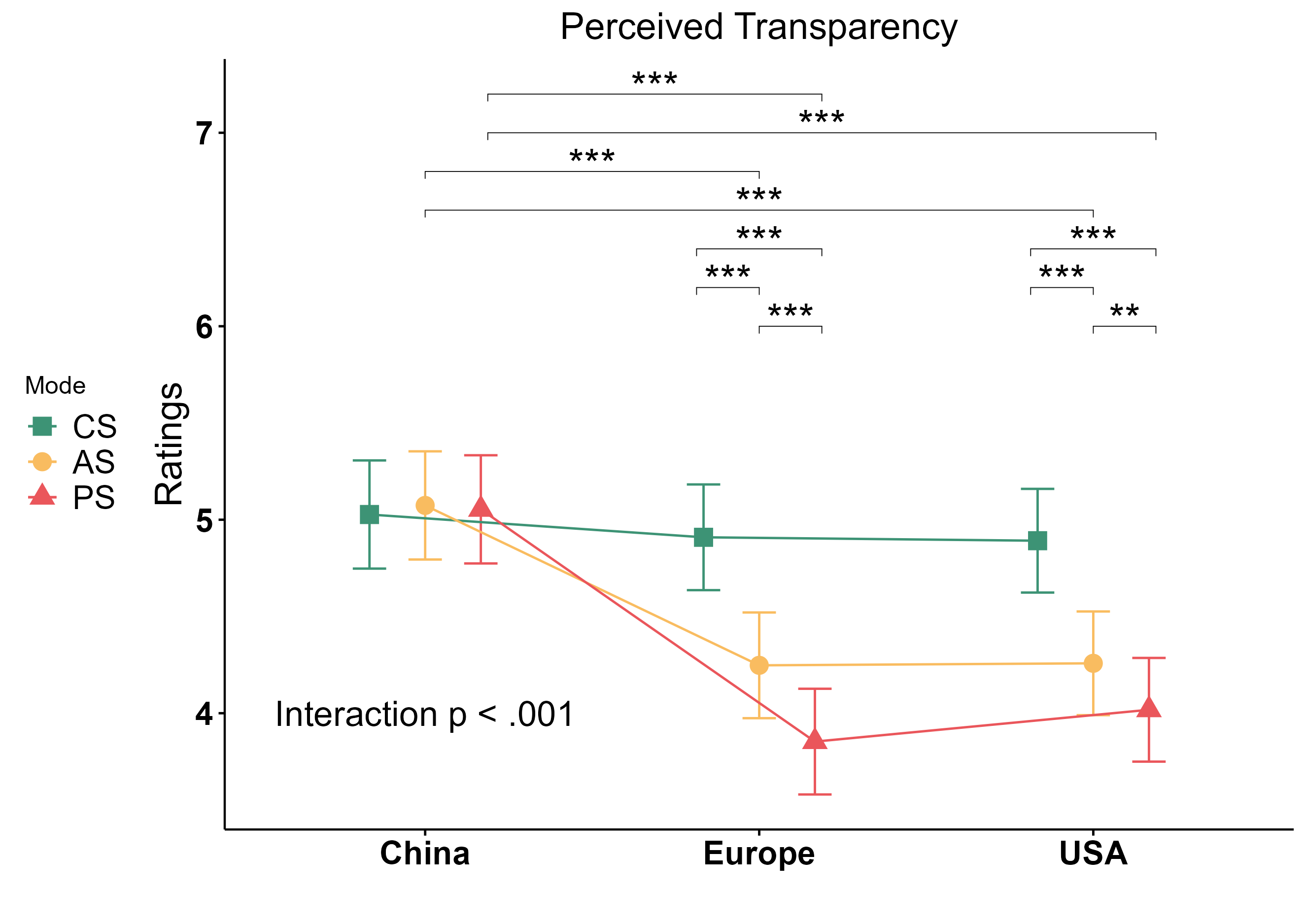} 
    \caption{Interaction plot of Perceived Transparency.} 
    \label{fig:TR}
\end{minipage}
\quad
\begin{minipage}{.45\linewidth}
  \centering
  \includegraphics[width=\linewidth]{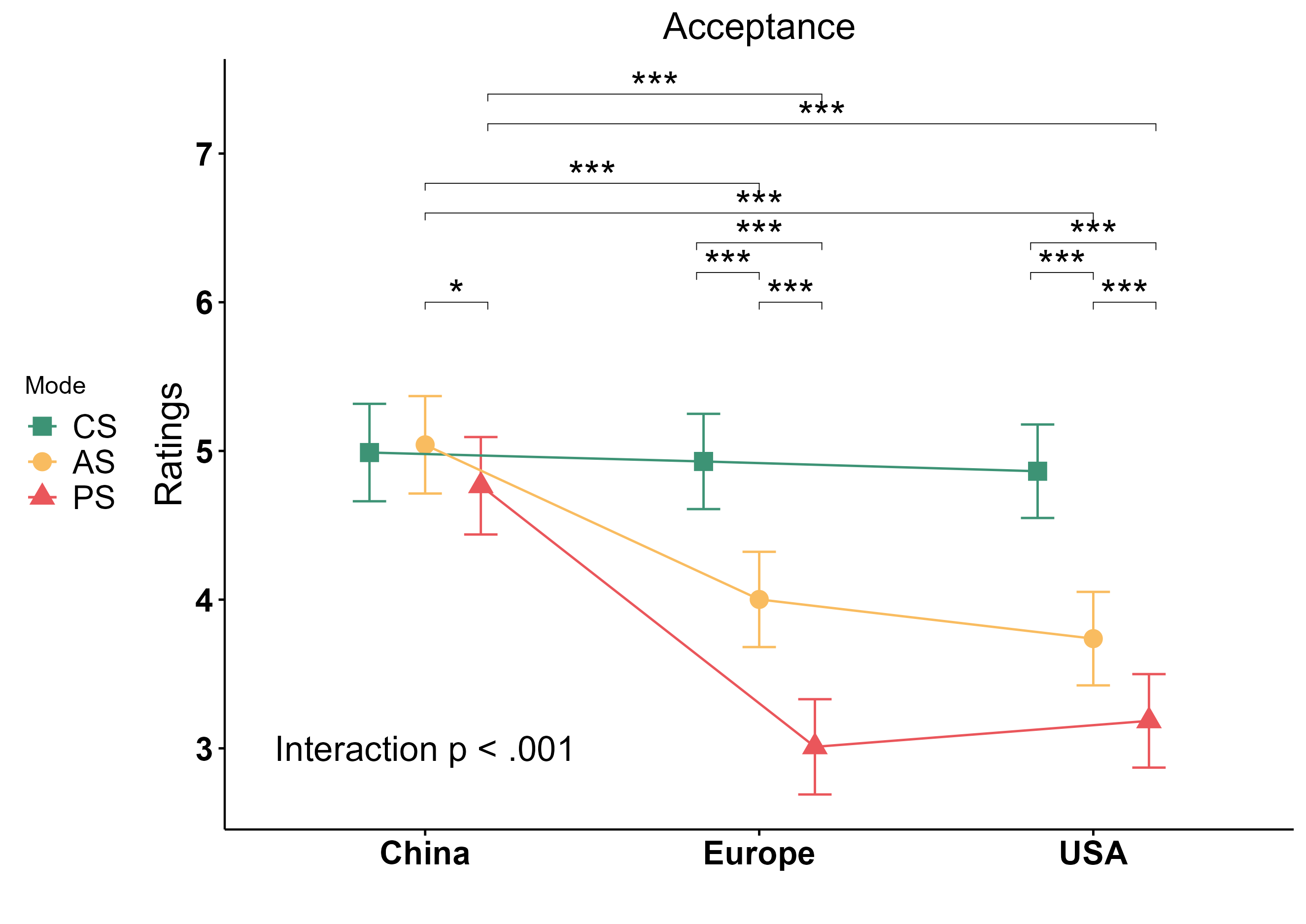} 
    \caption{Interaction plot of Acceptance.} 
    \label{fig:AC}
\end{minipage}

\end{figure*}}

\aptLtoX{\begin{table*}[hbt!]

    \centering 
  \caption{Estimated Marginal Means -- Perceived Capability}
\label{tab:eemPC}
\scalebox{0.9}{
\begin{tabular}{@{}l|cc|cc|cc@{}}
\toprule
 & \multicolumn{2}{c|}{China} & \multicolumn{2}{c|}{Europe} & \multicolumn{2}{c}{USA} \\ \midrule
 & \textbf{Mean} & \textbf{SE} & \textbf{Mean} & \textbf{SE} & \textbf{Mean} & \textbf{SE} \\ \midrule
CS & 4.96 & 0.14 & 4.49 & 0.14 & 4.58 & 0.14 \\
AS & 5.03 & 0.14 & 4.18 & 0.14 & 4.08 & 0.14 \\
PS & 4.89 & 0.14 & 3.74 & 0.14 & 3.76 & 0.14 \\ \bottomrule
\end{tabular}}
\end{table*}
\quad
\begin{table*}
  \centering
    \caption{Estimated Marginal Means -- Perceived Risk}
\label{tab:eemPR}
\scalebox{0.9}{
\begin{tabular}{@{}l|cc|cc|cc@{}}
\toprule
 & \multicolumn{2}{c|}{China} & \multicolumn{2}{c|}{Europe} & \multicolumn{2}{c}{USA} \\ \midrule
 & \textbf{Mean} & \textbf{SE} & \textbf{Mean} & \textbf{SE} & \textbf{Mean} & \textbf{SE} \\ \midrule
CS & 4.21 & 0.15 & 4.12 & 0.14 & 4.12 & 0.14 \\
AS & 4.27 & 0.15 & 4.88 & 0.14 & 4.98 & 0.14 \\
PS & 4.46 & 0.15 & 5.44 & 0.14 & 5.37 & 0.14 \\ \bottomrule
\end{tabular}}
\end{table*}

\bigskip

\begin{table*}
\centering
\caption{{\footnotesize Estimated Marginal Means -- Perceived Transparency}}
\label{tab:eemPT}
\scalebox{0.9}{
\begin{tabular}{@{}l|cc|cc|cc@{}}
\toprule
 & \multicolumn{2}{c|}{China} & \multicolumn{2}{c|}{Europe} & \multicolumn{2}{c}{USA} \\ \midrule
 & \textbf{Mean} & \textbf{SE} & \textbf{Mean} & \textbf{SE} & \textbf{Mean} & \textbf{SE} \\ \midrule
CS & 5.02 & 0.14 & 4.91 & 0.14 & 4.89 & 0.14 \\
AS & 5.07 & 0.14 & 4.25 & 0.14 & 4.26 & 0.14 \\
PS & 5.05 & 0.14 & 3.85 & 0.14 & 4.02 & 0.14 \\ \bottomrule
\end{tabular}}
\end{table*}
\quad
\begin{table*}
\centering
\caption{Estimated Marginal Means -- Acceptance}
\label{tab:eemAC}
\scalebox{0.9}{
\begin{tabular}{@{}l|cc|cc|cc@{}}
\toprule
 & \multicolumn{2}{c|}{China} & \multicolumn{2}{c|}{Europe} & \multicolumn{2}{c}{USA} \\ \midrule
 & \textbf{Mean} & \textbf{SE} & \textbf{Mean} & \textbf{SE} & \textbf{Mean} & \textbf{SE} \\ \midrule
CS & 4.99 & 0.17 & 4.93 & 0.16 & 4.86 & 0.16 \\
AS & 5.04 & 0.17 & 4.00 & 0.16 & 3.74 & 0.16 \\
PS & 4.77 & 0.17 & 3.01 & 0.16 & 3.18 & 0.16 \\ \bottomrule
\end{tabular}}

\end{table*}}{\begin{table*}[hbt!]
\centering
\begin{minipage}{.45\linewidth}
    \centering 
  \caption{Estimated Marginal Means -- Perceived Capability}
\label{tab:eemPC}
\scalebox{0.9}{
\begin{tabular}{@{}l|cc|cc|cc@{}}
\toprule
 & \multicolumn{2}{c|}{China} & \multicolumn{2}{c|}{Europe} & \multicolumn{2}{c}{USA} \\ \midrule
 & \textbf{Mean} & \textbf{SE} & \textbf{Mean} & \textbf{SE} & \textbf{Mean} & \textbf{SE} \\ \midrule
CS & 4.96 & 0.14 & 4.49 & 0.14 & 4.58 & 0.14 \\
AS & 5.03 & 0.14 & 4.18 & 0.14 & 4.08 & 0.14 \\
PS & 4.89 & 0.14 & 3.74 & 0.14 & 3.76 & 0.14 \\ \bottomrule
\end{tabular}}
\end{minipage}
\quad
\begin{minipage}{.45\linewidth}
  \centering
    \caption{Estimated Marginal Means -- Perceived Risk}
\label{tab:eemPR}
\scalebox{0.9}{
\begin{tabular}{@{}l|cc|cc|cc@{}}
\toprule
 & \multicolumn{2}{c|}{China} & \multicolumn{2}{c|}{Europe} & \multicolumn{2}{c}{USA} \\ \midrule
 & \textbf{Mean} & \textbf{SE} & \textbf{Mean} & \textbf{SE} & \textbf{Mean} & \textbf{SE} \\ \midrule
CS & 4.21 & 0.15 & 4.12 & 0.14 & 4.12 & 0.14 \\
AS & 4.27 & 0.15 & 4.88 & 0.14 & 4.98 & 0.14 \\
PS & 4.46 & 0.15 & 5.44 & 0.14 & 5.37 & 0.14 \\ \bottomrule
\end{tabular}}
\end{minipage}

\bigskip

\begin{minipage}{.45\linewidth}
\centering
\caption{{\footnotesize Estimated Marginal Means -- Perceived Transparency}}
\label{tab:eemPT}
\scalebox{0.9}{
\begin{tabular}{@{}l|cc|cc|cc@{}}
\toprule
 & \multicolumn{2}{c|}{China} & \multicolumn{2}{c|}{Europe} & \multicolumn{2}{c}{USA} \\ \midrule
 & \textbf{Mean} & \textbf{SE} & \textbf{Mean} & \textbf{SE} & \textbf{Mean} & \textbf{SE} \\ \midrule
CS & 5.02 & 0.14 & 4.91 & 0.14 & 4.89 & 0.14 \\
AS & 5.07 & 0.14 & 4.25 & 0.14 & 4.26 & 0.14 \\
PS & 5.05 & 0.14 & 3.85 & 0.14 & 4.02 & 0.14 \\ \bottomrule
\end{tabular}}
\end{minipage}
\quad
\begin{minipage}{.45\linewidth}
\centering
\caption{Estimated Marginal Means -- Acceptance}
\label{tab:eemAC}
\scalebox{0.9}{
\begin{tabular}{@{}l|cc|cc|cc@{}}
\toprule
 & \multicolumn{2}{c|}{China} & \multicolumn{2}{c|}{Europe} & \multicolumn{2}{c}{USA} \\ \midrule
 & \textbf{Mean} & \textbf{SE} & \textbf{Mean} & \textbf{SE} & \textbf{Mean} & \textbf{SE} \\ \midrule
CS & 4.99 & 0.17 & 4.93 & 0.16 & 4.86 & 0.16 \\
AS & 5.04 & 0.17 & 4.00 & 0.16 & 3.74 & 0.16 \\
PS & 4.77 & 0.17 & 3.01 & 0.16 & 3.18 & 0.16 \\ \bottomrule
\end{tabular}}
\end{minipage}

\end{table*}}

\subsection{Perceived Capability} 
\autoref{fig:PA} shows the predicted ratings (adjusted for age and gender) of perceived capability for three traffic surveillance modes across three regions. \sw{The main effects of region (F (2, 595) = 37.58, p < .001, $\eta^2_P$ = 0.11) and mode (F (2, 1196) = 77.58, p < .001, $\eta^2_P$ = 0.11) on perceived capability were both statistically significant. The interaction effect was also significant (F (4, 1196) = 15.29, p < .001, $\eta^2_P$ = 0.05). The effect of age was significant but extremely weak (F (1, 595) = 5.70, p = .017, $\eta^2_P$ = 0.009), the effect of gender was not significant (F (3, 594) = 1.05 , p = .351).} Simple effects analysis (see \autoref{tab:eemPC} for values of means and their standard errors [SE]) indicated the following:

\subsubsection{Regional comparisons under mode conditions} 
\sw{Chinese participants rated the perceived capability of all three modes significantly higher than both European participants (CS: p < .001, d = 0.62; AS: p < .001, d = 1.11; PS: p < .001, d = 1.50) and American participants (CS: p = .006, d = 0.51; AS: p < .001, d = 1.24; PS: p < .001, d = 1.48).} There were no significant differences between European and American participants in their ratings of the three surveillance modes. 

\subsubsection{Comparisons of modes within each region} 
\sw{Within Europe, participants rated CS significantly higher than both AS (p < .001, d = 0.40) and PS (p < .001, d = 0.98); and AS was rated significantly higher than PS (p < .001, d = 0.58). Likewise, within the USA, participants rated CS significantly higher than both AS (p = < .001, d = 0.65) and PS (p < .001, d = 1.07); and AS was rated significantly higher than PS (p < .001, d = 0.42).} In contrast, there were no significant differences in ratings among the three modes within China.

\subsection{Perceived Risk} 
\autoref{fig:PR} shows the ratings of perceived risk for three traffic surveillance modes across three regions. \sw{The main effects of region (F (2, 595) = 13.88, p < .001, $\eta^2_P$ = 0.04) and mode (F (2, 1196) = 179.28, p < .001, $\eta^2_P$ = 0.23) on perceived risk were both statistically significant. The interaction effect was also significant (F (4, 1196) = 25.75, p < .001, $\eta^2_P$ = 0.08). The effect of age (F (1, 595) = 0.67, p = .418) and gender (F (3, 595) = 0.38 , p = .686) were not significant.} Simple effects analysis (see \autoref{tab:eemPR} for values of means and their standard errors [SE]) indicated the following:

\subsubsection{Regional comparisons under mode conditions} 
\sw{Chinese participants rated the perceived risk of AS and PS significantly lower than both European participants (AS: p < .001, d = 0.71; PS: p < .001, d = 1.14) and American participants (AS: p < .001, d = 0.83; PS: p < .001, d = 1.05).} There were no significant differences between European and American participants in their ratings of the three surveillance modes. 

\subsubsection{Comparisons of modes within each region} 
\sw{Within Europe, participants rated CS significantly lower than both AS (p < .001, d = 0.88) and PS (p < .001, d = 1.52); and AS was rated significantly lower than PS (p < .001, d = 0.65). Likewise, within the USA, participants rated CS significantly lower than both AS (p < .001, d = 1.00) and PS (p < .001, d = 1.44); and AS was rated significantly lower than PS (p < .001, d = 0.45). Within China, participants rated PS significantly higher than CS (p = .012, d = 0.29).}

\subsection{Perceived Transparency} 
\autoref{fig:TR} shows the ratings of perceived transparency for three traffic surveillance modes across three regions. \sw{The main effects of region (F (2, 595) = 26.10, p < .001, $\eta^2_P$ = 0.08) and mode (F (2, 1196) = 120.46, p < .001, $\eta^2_P$ = 0.17) on perceived transparency were both statistically significant. The interaction effect was also significant (F (4, 1196) = 34.00, p < .001, $\eta^2_P$ = 0.10). The effect of age was significant but extremely weak (F (1, 595) = 4.54, p = .034, $\eta^2_P$ = 0.008), the effect of gender was not significant (F (3, 595) = 1.34 , p = .163).} Simple effects analysis (see \autoref{tab:eemPT} for values of means and their standard errors [SE]) indicated the following:

\subsubsection{Regional comparisons under mode conditions} 
\sw{Chinese participants rated the perceived transparency of AS and PS modes significantly higher than both European participants (AS: p < .001, d = 1.15; PS: p < .001, d = 1.67) and American participants (AS: p < .001, d = 1.13; PS: p < .001, d = 1.44).} There were no significant differences between European and American participants in their ratings of the three surveillance modes.

\subsubsection{Comparisons of modes within each region} 
\sw{Within Europe, participants rated CS significantly higher than both AS (p < .001, d = 0.92) and PS (p < .001, d = 1.47); and AS was rated significantly higher than PS (p < .001, d = 0.55). Likewise, within the USA, participants rated CS significantly higher than both AS (p < .001, d = 0.88) and PS (p < .001, d = 1.21); and AS was rated significantly higher than PS (p = .002, d = 0.33).} In contrast, there were no significant differences in ratings among the three modes within China.

\subsection{Acceptance} 
\autoref{fig:AC} shows the ratings of acceptance for three traffic surveillance modes across three regions. \sw{The main effects of region (F (2, 595) = 39.46, p < .001, $\eta^2_P$ = 0.12) and mode (F (2, 1196) = 244.41, p < .001, $\eta^2_P$ = 0.29) on acceptance were both statistically significant. The interaction effect was also significant (F (4, 1196) = 45.47, p < .001, $\eta^2_P$ = 0.13). The effect of age (F (1, 595) = 3.20, p = .074) and gender (F (3, 595) = 0.32 , p = .723) were not significant.} Simple effects analysis (see \autoref{tab:eemAC} for values of means and their standard errors [SE]) indicated the following:

\subsubsection{Regional comparisons under mode conditions} 
\sw{Chinese participants rated the acceptance of AS and PS modes significantly higher than both European participants (AS: p < .001, d = 1.04; PS: p < .001, d = 1.76) and American participants (AS: p < .001, d = 1.30; PS: p < .001, d = 1.58).}

\subsubsection{Comparisons of modes within each region}
\sw{Within Europe, participants rated CS significantly higher than both AS (p < .001, d = 0.93) and PS (p < .001, d = 1.92); and AS was rated significantly higher than PS (p < .001, d = 0.99). Likewise, within the USA, participants rated CS significantly higher than both AS (p < .001, d = 1.13) and PS (p < .001, d = 1.68); and AS was rated significantly higher than PS (p < .001, d = 0.55). Within China, participants rated AS significantly higher than PS ( p = .017, d = 0.28).}

{\color{black}\subsection{Summary of Key Findings}

In the following, we summarize the key findings and discuss the causes and implications in the next section.

For conventional surveillance (CS), there are no regional differences in perceived risk, transparency, or acceptance. Overall, AI-powered surveillance (AS and PS) was rated less acceptable with lower capability, higher risk and lower transparency than conventional surveillance (CS). Across all cultures, public-shaming surveillance (PS) is perceived as the riskiest and least acceptable mode.

Chinese participants reported smaller differences between the three surveillance modes, whereas European and American participants showed more pronounced distinctions between the modes. There are no significant differences between Europe and the USA, except that AS is slightly more accepted in Europe than in the USA. AI-powered surveillance (AS and PS) is viewed as more capable, less risky, more transparent, and more accepted in China compared to Europe or the USA.

}

\section{Discussion}


This study advances research on surveillance technology and human–AI interaction by systematically examining how people across different cultural contexts perceive traffic surveillance systems. While prior work has often treated surveillance acceptance as either a technical design challenge (e.g., transparency, privacy-preserving methods) \cite{TAM_VS, chi24drone} or a cultural question \cite{golda2022perception, Guardian2015, AIopti} in isolation, our findings reveal the interplay between both. By comparing conventional surveillance (CS), AI-enhanced surveillance (AS), and AI-enhanced surveillance with public shaming (PS) across China, Europe, and the United States, we provide a nuanced view of how AI reshapes public perceptions of road monitoring and how cultural contexts mediate these changes.

\subsection{\textbf{Convergent Views on Conventional Surveillance}}

Much of the HCI literature assumes that cultural differences are the dominant factor in shaping attitudes toward surveillance \cite{Guardian2015, Yang_2024}. Yet our findings show that in the case of conventional surveillance, attitudes converge across regions. Despite China’s more expansive surveillance infrastructure, participants in China, Europe, and the USA shared strikingly similar views on risk, transparency, and acceptance. Although the perceived capability of conventional traffic surveillance (CS) was rated significantly higher by Chinese participants than European and American participants, the magnitude of this difference was noticeably smaller than that observed for the AI-enhanced systems (AS and PS), as shown in \autoref{fig:PA}, which also indicate a convergent view.

This convergence challenges the tendency in prior research to overstate East–West divides and treat cultural context as a catch-all explanation \cite{NYTimes2018, Economist2023}. Instead, our findings suggest that familiarity and standardization may play a more powerful role than cultural orientation in shaping attitudes toward long-standing technologies. For HCI, this calls for a recalibration: rather than assuming cultural difference always drives divergence, we must examine how global infrastructures and decades of exposure can create shared baselines of acceptance.

\subsection{\textbf{AI Isn’t Automatically Perceived as an Upgrade}}

Previous research on “smart” surveillance often emphasizes the promise of AI to improve efficiency, accuracy, and safety \cite{Adewopo2023, sharma2024video}. Yet our results show that these benefits are not recognized by the public: European and American participants rated AI surveillance as less capable, more risky, and less transparent than conventional ones, while even Chinese participants did not see AI as more significantly capable than the conventional,  as shown in \autoref{fig:PA}.

This challenges the techno-optimism prevalent, which often take for granted that AI is perceived as progress \cite{AIopti}. One of the key reasons to implement AI in traffic surveillance is its potential to significantly enhance security through real-time threat detection and response, as well as to offer improved accuracy while minimizing human error~\cite{forbes2023intelligent}. By centering \rev{primarily} on technical capacity, much prior work \rev{overlooks} the gap between what AI is supposed to do and what people actually believe it can do. Our findings \rev{suggest that technical performance of} AI systems \rev{alone may not guarantee greater public preference or acceptance. Complementary efforts, such as} improving AI literacy, communicating system benefits, and designing for explainability\rev{, warrant further investigation as potential approaches to addressing this gap.}

\subsection{\textbf{Public Shaming \rev{Is} an \rev{Unfavorable} Design Strategy}}

Our findings further demonstrate that public shaming (PS) is widely \rev{viewed unfavorably}. \rev{Beyond the lower preference for} AI-enhanced traffic surveillance \rev{relative to} conventional surveillance, significant differences were observed between the two AI-enhanced modes studied (AS and PS). European and American participants rated PS as the least preferred across all measures. Although Chinese participants \rev{showed smaller} differences between AS and PS, they still perceived PS as significantly higher in risk and lower in acceptance than AS\rev{,  while rating its} capability \rev{as comparable}.

While PS could theoretically serve as a deterrent, participants perceived it as riskier, less acceptable, and no more capable than ordinary AI-enhanced surveillance. The divergence between potential effectiveness and perceived legitimacy highlights a core HCI insight: systems must align with human values, not merely outcomes. This stands in contrast to surveillance practices that emphasize compliance over dignity, such as China’s jaywalker-shaming campaigns \cite{Grossman2018ChineseFacial}. Our data suggest that such coercive mechanisms erode \rev{acceptance} and increase perceived risks, even when they promise effectiveness. A human-centered approach should instead emphasize positive reinforcement, fairness, and respect for dignity. \rev{Our findings suggest that caution may be warranted when considering PS} in regulatory frameworks, not only for privacy \rev{concerns} but also because public rejection \rev{may limit} its long-term viability.

\subsection{\textbf{Cultural Patterns of Acceptance}}

While conventional surveillance elicited convergence, AI-enhanced systems revealed divergence. Chinese participants were more accepting of AI surveillance overall, while Europeans and Americans showed sharper distinctions, favoring CS and rejecting PS. These findings resonate with cultural theory: in collectivist contexts, surveillance may be normalized as a collective good, while individualist contexts place stronger emphasis on privacy and autonomy \cite{BondSmith1996, Kwan2023}.

Related work has emphasized these cultural divides in abstract terms, but our findings situate them within a concrete infrastructural domain: traffic surveillance. \rev{One possible explanation is differences in} normalization and exposure. \rev{In China, the widespread} integration of AI into daily life—from payments to policing—\rev{may contribute to} higher acceptance. In contrast, Western skepticism \rev{may reflect} not only regulatory debates but also cultural \rev{narratives} of “AI rebellion” \cite{coman2017social} and “AI displacement” \cite{maas2019international}. \rev{Although these mechanisms were not directly examined in this study, they provide a plausible context for interpreting the observed differences.}

\rev{The findings suggest that} one-size-fits-all approaches are inadequate. \rev{Instead,} systems \rev{may benefit from adapting} to regional expectations—\rev{for example,} emphasizing efficiency in contexts such as China \rev{while} prioritizing transparency and accountability in Europe and the U.S. \rev{More broadly,} international governance frameworks \rev{may need to} balance global principles\rev{, such as} fairness, transparency, \rev{and} accountability, with \rev{sensitivity to local} cultural \rev{contexts}.

\section{Limitations and Future Work}

Although our study includes three major regions—namely the USA, China, and Europe—these do not fully represent the diversity of smaller countries or developing regions. Future work should extend the analysis to a broader set of cultural and socioeconomic contexts to provide a more inclusive understanding of perceptions of AI-powered traffic surveillance.

We also did not examine intra-regional variation. Europe comprises multiple countries with distinct traffic regulations, while the USA and China contain states and provinces with differing legal frameworks and traffic practices. Additionally, large cities and small towns differ in infrastructure, population density, and social visibility, which may influence both technology acceptance and privacy concerns. For example, public shaming may have stronger social consequences in small communities, whereas large urban displays may be perceived as more intimidating due to scale. Future studies should incorporate such intra-regional factors to provide a more nuanced analysis.

Our survey described AI-enabled surveillance at a high level. More specific scenarios—such as varying degrees of automation, hybrid human–AI decision-making, or preventive versus post-hoc monitoring—may influence perceptions and should be explored in future work.

\rev{This study measured participants' stated perceptions of road surveillance systems using static picture-and-text vignettes rather than observing behavior in real-world settings \cite{ZWang_PhD}. While this approach enables systematic comparison of surveillance system types under controlled conditions, it necessarily simplifies the situational detail of an actual encounter with these systems, such as what the system detects, what information is displayed, to whom, and what consequences follow. A video-based or more situationally specific vignette could convey this detail more concretely and may therefore yield different, and potentially more ecologically valid, perceptions than those captured here.} 

\rev{Moreover, stated attitudes do not necessarily translate into actual behavior: prior research on the ``privacy paradox'' has shown that people often express concerns about surveillance yet behave differently when encountering such systems in practice \cite{privacy_paradox}. Factors such as habituation, perceived necessity, and situational context may therefore lead to behavioral responses that differ from the attitudes reported here. Accordingly, our findings should be interpreted as comparative evidence of perceived acceptability across surveillance systems and regions, rather than predictions of real-world human behaviors. Future work should build on this exploratory study using richer situational stimuli (e.g., video-based scenarios depicting concrete road situations) and validate these findings through field studies, longitudinal deployments, or behavioral trace analysis (e.g., avoidance behavior, opt-out or complaint rates) to examine whether the observed regional and system-level differences persist in practice.}

\rev{Another limitation is that the PS condition combined multiple features—facial recognition, identification and disclosure of personal information, and public display—rather than manipulating them independently. Consequently, we cannot attribute the lower acceptance of PS to any single feature, or determine whether it resulted from their combined effects. Future research should adopt a factorial design that independently varies these features to disentangle their individual and interactive contributions to human perceptions.}

Due to institutional data minimization requirements, we did not collect participants’ educational background. However, AI literacy and education may shape attitudes toward automated surveillance and should be considered in future studies.

Finally, this work relied on quantitative methods. Complementary qualitative approaches, such as interviews or focus groups, could provide deeper insights into pedestrian perceptions, concerns, and expectations regarding AI-powered traffic surveillance.

\section{Conclusions} 
Given the global growth in traffic surveillance and recent proposals or pilots supplementing such systems with AI analysis and potentially public shaming, we find it timely to conduct a cross-cultural survey in China, Europe, and the USA to understand road users' perspectives. Regarding conventional road surveillance (cameras only), respondents across regions show comparable assessments of risk, transparency, and acceptance, possibly due to its longstanding presence in all three regions. The addition of AI-enhancement lowers rankings across all scales, including perceived capability, for Europeans and Americans only, while Chinese participants seem to consider the technologies as being more comparable. Yet all three groups similarly ranked AI surveillance with public shaming lower than the other two alternatives. Thus we see significant similarities as well as differences in perspectives across regions. Road users' familiarity with the technology, regional culture and media traditions, alignment with values, trust in AI, and perceptions of benefits may all shape perspectives and contribute to acceptance. For now, people out and about in public are sensitive to having AI eyes on them, even when their role is in support of overall safety.

\begin{acks}
\rev{We thank all anonymous reviewers for their efforts and valuable inputs. SY expresses his sincere gratitude to Cecilia Jakobsson Bergstad for her guidance on the methodology. We acknowledge the Wallenberg AI, Autonomous Systems and Software Program -- Humanity and Society (WASP-HS). This research was primarily funded by the Marianne and Marcus Wallenberg Foundation.}

\end{acks}

\bibliographystyle{ACM-Reference-Format}
\bibliography{sample-base}

\onecolumn
\appendix
\section{Appendix}

\begin{table*}[h!]
\centering
\caption{Reliability values of the four  question items: PC, PR, PT, and AC.}
\label{table:reliability_values}
\begin{tabular}{llccc}
\toprule
\textbf{Question items} & \textbf{95\% CI} & \textbf{Cronbach's $\alpha$} & \textbf{Guttman's $\lambda$} \\
\midrule

\textbf{Perceived Capability (PC)} & Feldt: [0.87, 0.89] & 0.88 & 0.86 \\
 & Duhachek: [0.87, 0.89] & & \\
  & &   &   \\

\textbf{Perceived Risk (PR)} & Feldt: [0.88, 0.90] & 0.89 & 0.86 \\
 & Duhachek: [0.88, 0.90] & & \\
  & &   &   \\

\textbf{Perceived Transparency (PT)} & Feldt: [0.86, 0.88] & 0.87 & 0.84 \\
 & Duhachek: [0.86, 0.88] & & \\
  & &   &   \\

\textbf{Acceptance (AC)} & Feldt: [0.94, 0.95] & 0.95 & 0.93 \\
 & Duhachek: [0.94, 0.95] & & \\

\bottomrule
\end{tabular}
\end{table*}

\aptLtoX{\begin{figure*}[hbt!]
  \centering
    \includegraphics[width=\linewidth]{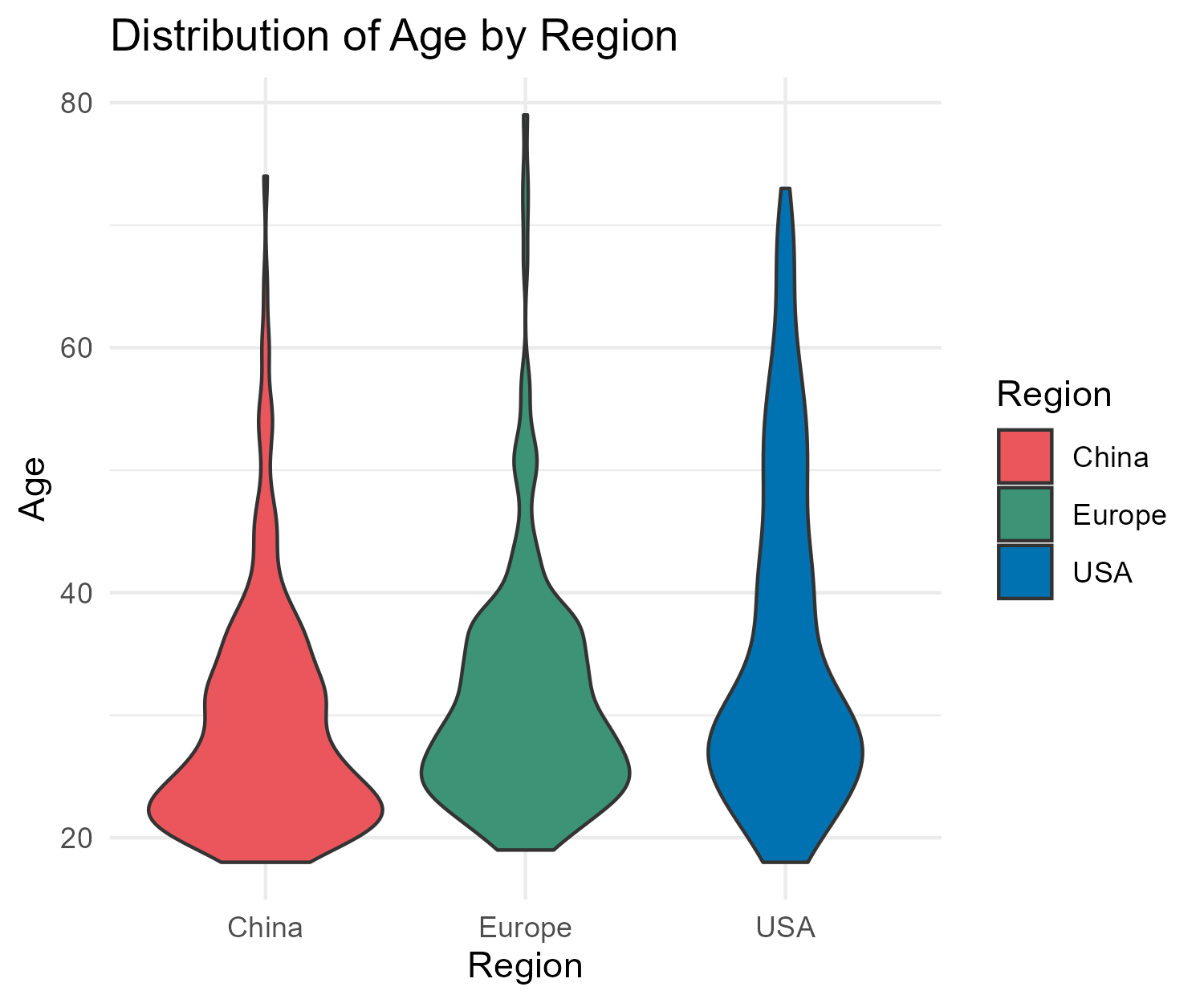} 
    \caption{The distribution of age by region.} 
    \label{fig:age}
\end{figure*}
\quad
\begin{figure*}
   \centering
    \includegraphics[width=\linewidth]{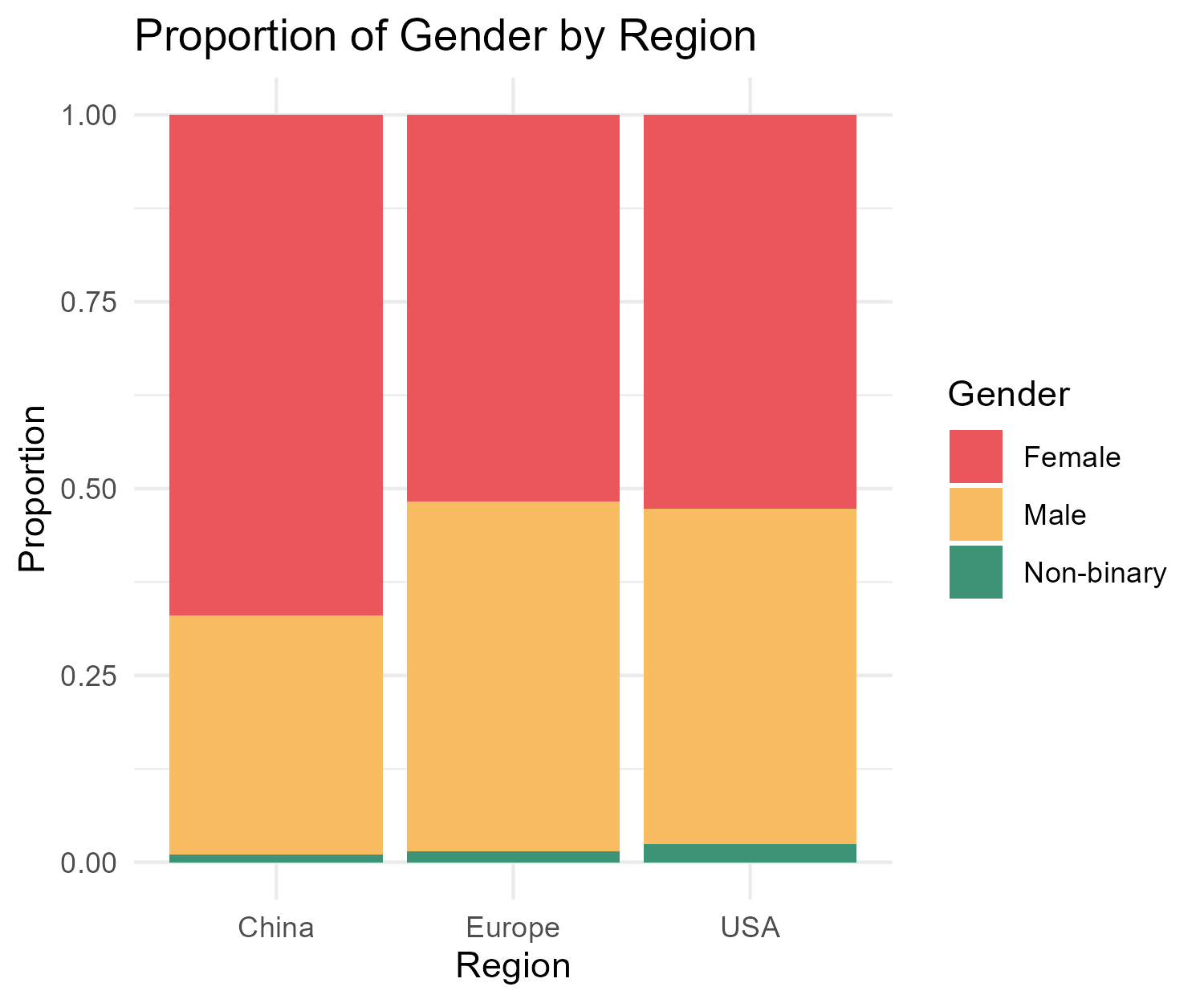}
    \caption{The distribution of gender by region.} 
    \label{fig:gender}
\end{figure*}}{\begin{figure*}[hbt!]
\centering
\begin{minipage}{.4\linewidth}
  \centering
    \includegraphics[width=\linewidth]{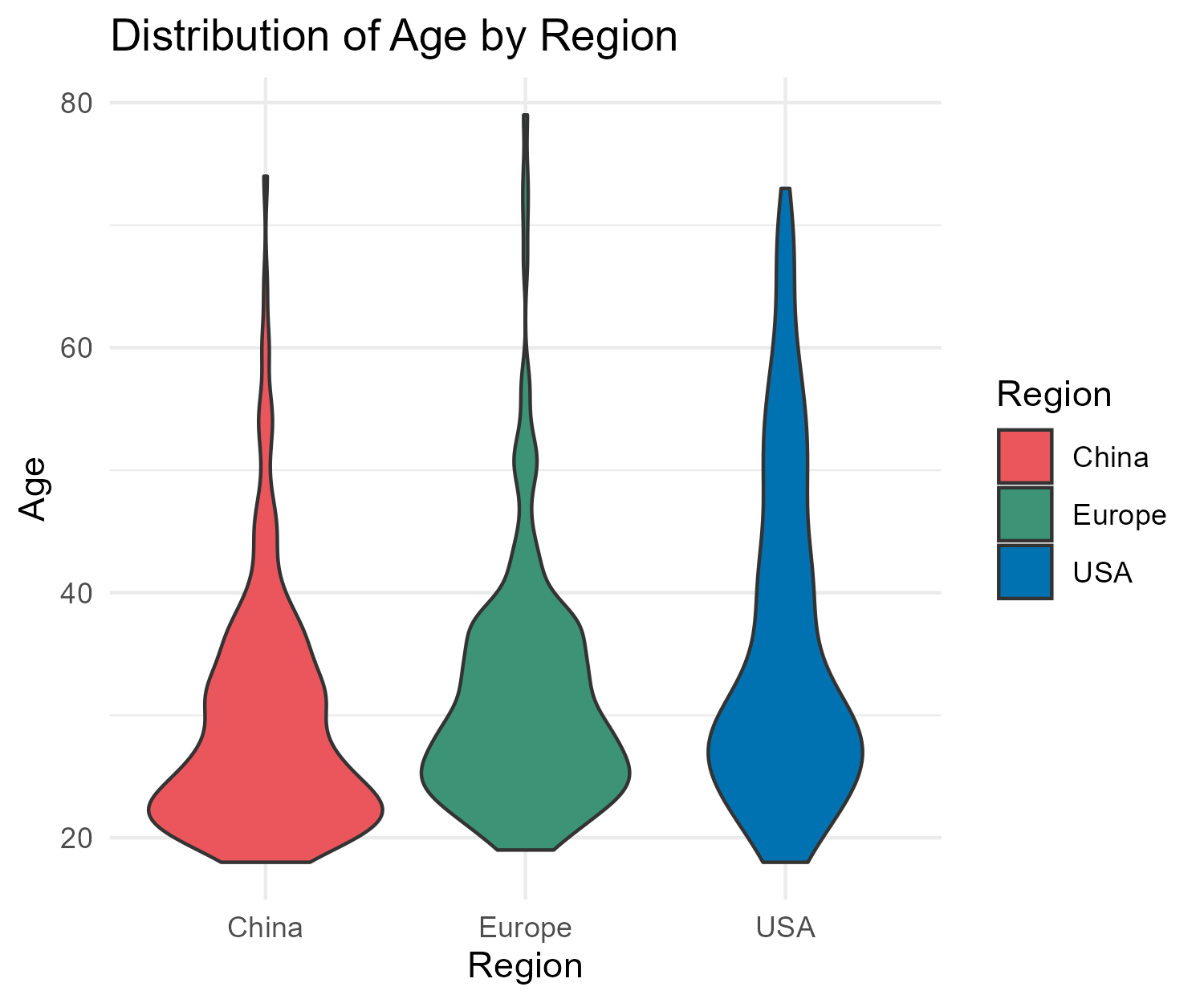} 
    \caption{The distribution of age by region.} 
    \label{fig:age}
\end{minipage}
\quad
\begin{minipage}{.35\linewidth}
   \centering
    \includegraphics[width=\linewidth]{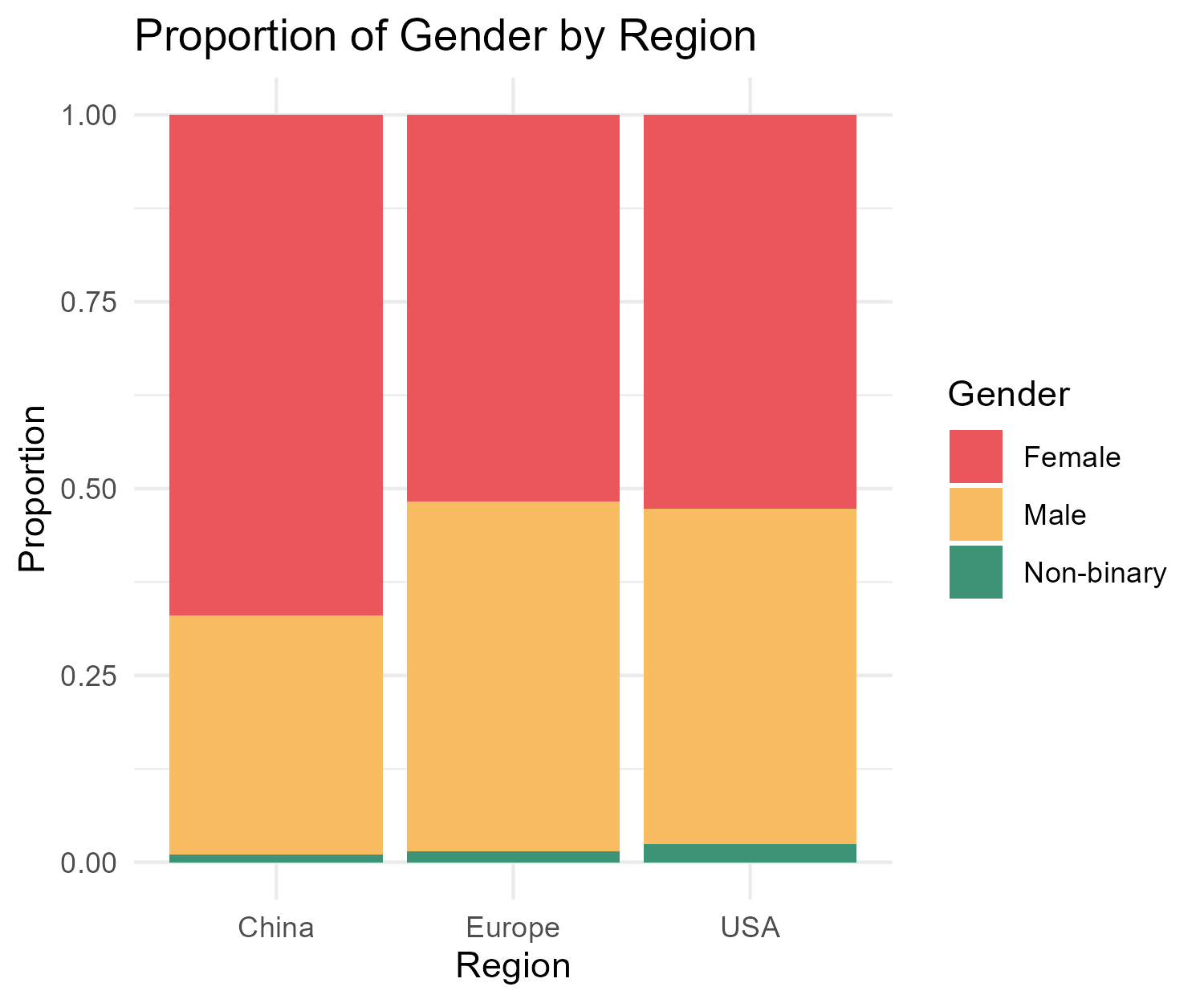}
    \caption{The distribution of gender by region.} 
    \label{fig:gender}
\end{minipage}
\end{figure*}}

\end{document}